\documentclass[journal]{IEEEtran}
\makeatletter

\usepackage{blindtext}
\usepackage[T1]{fontenc}
\usepackage{textcomp}
\usepackage{amsmath}
\usepackage{amssymb}
\usepackage{bm}
\usepackage{mdwmath}
\usepackage{cases}
\usepackage{graphicx}
\usepackage{array}

\graphicspath{{Figure/PDF/}{Biography/PDF/}}
\usepackage[dvipsnames]{xcolor}
\definecolor{myblue}{rgb}{0,0.4980,1} % Azure
\definecolor{myred}{rgb}{0.8706,0.1608,0.0627} % Chinese red
\usepackage[caption=false,font=footnotesize,subrefformat=parens]{subfig}
\usepackage[nosort,noadjust]{cite}
\usepackage{url}
\usepackage{tabularray}
\UseTblrLibrary{varwidth}
\UseTblrLibrary{diagbox}
\ExplSyntaxOn
\tl_const:Nn \c__tblr_valign_M_tl {M}
\tl_new:N    \g__tblr_cell_valign_sub_tl

\keys_define:nn { tblr-column }
  {
    M .meta:n = { valign = M },
  }

\cs_gset_protected:Npn \__tblr_get_cell_alignments:nn #1 #2
  {
    \group_begin:
    \tl_gset:Nx \g__tblr_cell_halign_tl
      { \__tblr_data_item:neen { cell } {#1} {#2} { halign } }
    \tl_set:Nx \l__tblr_v_tl
      { \__tblr_data_item:neen { cell } {#1} {#2} { valign } }
    \tl_case:NnF \l__tblr_v_tl
      {
        \c__tblr_valign_t_tl
          {
            \tl_gset:Nn \g__tblr_cell_valign_tl {m}
            \tl_gset:Nn \g__tblr_cell_middle_tl {t}
            \tl_gclear:N \g__tblr_cell_valign_sub_tl
          }
        \c__tblr_valign_m_tl
          {
            \tl_gset:Nn \g__tblr_cell_valign_tl {m}
            \tl_gset:Nn \g__tblr_cell_middle_tl {m}
            \tl_gclear:N \g__tblr_cell_valign_sub_tl
          }
        \c__tblr_valign_M_tl
          {
            \tl_gset:Nn \g__tblr_cell_valign_tl {m}
            \tl_gset:Nn \g__tblr_cell_valign_sub_tl {M}
            \tl_gset:Nn \g__tblr_cell_middle_tl {t}
          }
        \c__tblr_valign_b_tl
          {
            \tl_gset:Nn \g__tblr_cell_valign_tl {m}
            \tl_gset:Nn \g__tblr_cell_middle_tl {b}
            \tl_gclear:N \g__tblr_cell_valign_sub_tl
          }
      }
      {
        \tl_gset_eq:NN \g__tblr_cell_valign_tl \l__tblr_v_tl
        \tl_gclear:N \g__tblr_cell_middle_tl
      }
    \group_end:
  }

\cs_gset_protected:Npn \__tblr_build_cell_content:NN #1 #2
  {
    \hbox_set_to_wd:Nnn \l__tblr_a_box { \l__tblr_cell_wd_dim }
      {
        \tl_if_eq:NnTF \g__tblr_cell_halign_tl {j}
          { \__tblr_get_cell_text:nn {#1} {#2} \hfil }
          {
            \tl_if_eq:NnF \g__tblr_cell_halign_tl {l} { \hfil }
            \__tblr_get_cell_text:nn {#1} {#2}
            \tl_if_eq:NnF \g__tblr_cell_halign_tl {r} { \hfil }
          }
      }
    \vbox_set_to_ht:Nnn \l__tblr_b_box { \l__tblr_cell_ht_dim }
      {
        \tl_case:Nn \g__tblr_cell_valign_tl
          {
            \c__tblr_valign_m_tl
              {
                \vfil
                \bool_lazy_and:nnT
                  { \int_compare_p:nNn { \l__tblr_cell_rowspan_tl } < {2} }
                  { !\tl_if_eq_p:NN \c__tblr_valign_M_tl \g__tblr_cell_valign_sub_tl }
                  {
                    \box_set_ht:Nn \l__tblr_a_box
                      { \__tblr_data_item:nen { row } {#1} { @row-upper } }
                    \box_set_dp:Nn \l__tblr_a_box
                      { \__tblr_data_item:nen { row } {#1} { @row-lower } }
                  }
                \box_use:N \l__tblr_a_box
                \vfil
              }
            \c__tblr_valign_h_tl
              {
                \box_set_ht:Nn \l__tblr_a_box
                  { \__tblr_data_item:nen { row } {#1} { @row-head } }
                \box_use:N \l__tblr_a_box
                \vfil
              }
            \c__tblr_valign_f_tl
              {
                \vfil
                \int_compare:nNnTF { \l__tblr_cell_rowspan_tl } < {2}
                  {
                    \box_set_dp:Nn \l__tblr_a_box
                      { \__tblr_data_item:nen { row } {#1} { @row-foot } }
                  }
                  {
                    \box_set_dp:Nn \l__tblr_a_box
                      {
                        \__tblr_data_item:nen
                          { row }
                          { \int_eval:n { #1 + \l__tblr_cell_rowspan_tl - 1 } }
                          { @row-foot }
                      }
                  }
                \box_use:N \l__tblr_a_box
              }
          }
        \hrule height ~ 0pt %% zero depth
      }
    \vbox_set_to_ht:Nnn \l__tblr_c_box
      { \l__tblr_row_ht_dim - \l__tblr_row_abovesep_dim }
      {
        \box_use:N \l__tblr_b_box
        \vss
      }
    \skip_horizontal:n { \l__tblr_x_tl }
    \box_use:N \l__tblr_c_box
    \skip_horizontal:n { \l__tblr_y_tl - \l__tblr_cell_wd_dim + \l__tblr_w_tl }
  }
\ExplSyntaxOff

\newcommand{\colorhypersetup}{\@ifpackageloaded{hyperref}{\hypersetup{%
	bookmarksopen=true,%
	bookmarksnumbered=true,%
	pdfpagemode={UseOutlines},%default
	pdfstartview={FitH},%
	colorlinks=true,%
	linkcolor={myred},%
	citecolor={orange}
}}{\empty}}
\newcommand{\blackhypersetup}{\@ifpackageloaded{hyperref}{\hypersetup{%
	bookmarksopen=true,%
	bookmarksnumbered=true,%
	pdfpagemode={UseOutlines},%default
	pdfstartview={FitH},%
	colorlinks=true,%
	allcolors={black}
}}{\empty}}
\colorhypersetup
\usepackage{mfirstuc-english}
\MFUhyphentrue
\usepackage[version3,description,IEEEtran]{eacro}
\acsetup{%
    pages/display=all,%
    pages/seq/use=false,%
    pages/name=true,%
    pages/fill={\quad},%
    make-links=false,%
    case-sensitive=false,
}
\DeclareAcronym{iot}{
	short = IoT,
	long = Internet of things}
\DeclareAcronym{ai}{
    short = AI,
    long = artificial intelligence}
\DeclareAcronym{lora}{
    short = LoRa,
    long = long range}
\DeclareAcronym{2g}{
    short = 2G,
    long = second generation}
\DeclareAcronym{3g}{
    short = 3G,
    long = third generation}
\DeclareAcronym{4g}{
    short = 4G,
    long = fourth generation}
\DeclareAcronym{5g}{
    short = 5G,
    long = fifth generation}
\DeclareAcronym{6g}{
    short = 6G,
    long = sixth generation}
\DeclareAcronym{nbiot}{
    short = NB-IoT,
    long = narrowband \acs*{iot}}
\DeclareAcronym{aiot}{
    short = A-IoT,
    long = ambient \acs*{iot}}
\DeclareAcronym{3gpp}{
    short = 3GPP,
    long = 3rd Generation Partnership Project}
\DeclareAcronym{ruc}{
    short = rUC,
    long = representative use case}
\DeclareAcronym{ran}{
    short = RAN,
    long = radio access network}
\DeclareAcronym{bs}{
    short = BS,
    long = base station}
\DeclareAcronym{ue}{
    short = UE,
    long = user equipment}
\DeclareAcronym{ask}{
    short = ASK,
    long = Amplitude-Shift Keying}
\DeclareAcronym{fsk}{
    short = FSK,
    long = frequency-shift keying}
\DeclareAcronym{psk}{
    short = PSK,
    long = Phase-Shift Keying}
\DeclareAcronym{ook}{
    short = OOK,
    long = On-Off Keying}
\DeclareAcronym{qam}{
    short = QAM,
    long = Quadrature Amplitude Modulation}
\DeclareAcronym{bpsk}{
    short = BPSK,
    long = Binary \acs*{psk}}
\DeclareAcronym{rf}{
    short = RF,
    long = Radio Frequency}
\DeclareAcronym{rfid}{
    short = RFID,
    long = Radio Frequency Identification}
\DeclareAcronym{dpi}{
    short = DPI,
    long = Direct Path Interference}
\DeclareAcronym{sic}{
    short = SIC,
    long = Successive Interference Cancellation}
\DeclareAcronym{csi}{
    short = CSI,
    long = Channel State Information}
\DeclareAcronym{snr}{
    short = SNR,
    long = Signal-to-Noise Ratio}
\DeclareAcronym{tdma}{
    short = TDMA,
    long = Time-Division Multiple Access}
\DeclareAcronym{fdma}{
    short = FDMA,
    long = Frequency-Division Multiple Access}
\DeclareAcronym{usrp}{
    short = USRP,
    long = Universal Software Radio Peripheral}
\DeclareAcronym{arm}{
    short = ARM,
    long = advanced RISC machines}
\DeclareAcronym{ber}{
    short = BER,
    long = Bit Error Rate}
\DeclareAcronym{stc}{
    short = STC,
    long = Space-Time Code}
\DeclareAcronym{stbc}{
    short = STBC,
    long = Space-Time Block Code}
\DeclareAcronym{noma}{
    short = NOMA,
    long = Non-Orthogonal Multiple Access}
\DeclareAcronym{cbma}{
    short = CBMA,
    long = Coded-Backscatter Multiple Access}
\DeclareAcronym{crc}{
    short = CRC,
    long = Cyclic Redundancy Check}

\usepackage{algpseudocode}
\newcounter{MYalgorithmic}

{%
	\vspace*{3pt}
	\hrule height 1pt}
\newcounter{MYitem}[MYalgorithmic]

\newcommand{\MYlabel}[1]{\def\@currentlabel{\theALG@line}\label{#1}}
\newcommand{\upperroman}[1]{\uppercase\expandafter{\romannumeral#1}}

\newcommand{\myunit}[1]{%
	\ifmmode
		\mathrm{#1}
	\else
		$ \mathrm{#1} $% <-this % stops a space
	\fi}
\newcommand{\murm}{%
	\ifmmode
		\text{\textmu}
	\else
		\textmu
	\fi}

\newlength{\mysinglefigwidth}
\newlength{\mymultifigwidth}
\ifCLASSOPTIONonecolumn
	\AtBeginDocument{\setlength{\mysinglefigwidth}{0.8\linewidth}}
\else
	\AtBeginDocument{\setlength{\mysinglefigwidth}{\linewidth}}
\fi
\ifCLASSOPTIONonecolumn
	\AtBeginDocument{\setlength{\mymultifigwidth}{0.5\linewidth}}
\else
	\AtBeginDocument{\setlength{\mymultifigwidth}{\linewidth}}
\fi

\usepackage{soul}
\definecolor{acrobatyellow}{rgb}{1,0.8196,0}
\colorlet{highlightyellow}{acrobatyellow!30!white}
\sethlcolor{highlightyellow}

\makeatother
\begin{document}
%% *************************************************************************
\title{Ambient IoT towards 6G: Standardization, Potentials, and Challenges}

\author{Kan~Zheng,~\IEEEmembership{Fellow,~IEEE}, Rongtao~Xu, Jie~Mei,~\IEEEmembership{Member,~IEEE}, Haojun~Yang,~\IEEEmembership{Member,~IEEE}, Lei~Lei,~\IEEEmembership{Senior~Member,~IEEE}, and~Xianbin~Wang,~\IEEEmembership{Fellow,~IEEE}}

\maketitle

\begin{abstract}
The Ambient Internet of Things (A-IoT) has emerged as a critical direction for achieving effective connectivity as the IoT system evolves to 6G. However, the introduction of A-IoT technologies, particularly involving backscatter modulation, poses numerous challenges for system design and network operations. This paper surveys current standardization efforts, highlights potential challenges, and explores future directions for A-IoT. It begins with a comprehensive overview of ongoing standardization initiatives by the 3rd Generation Partnership Project (3GPP) on A-IoT, providing a solid foundation for further technical research in both industry and academia. Building upon this groundwork, the paper conducts an analysis of critical enabling technologies for A-IoT. Moreover, a comprehensive A-IoT demonstration system is designed to showcase the practical viability and efficiency of A-IoT techniques, supported by field experiments. We finally address ongoing challenges associated with A-IoT technologies, providing valuable insights for future research endeavors.
\end{abstract}

\begin{IEEEkeywords}
Internet of things (IoT), backscatter communications, and 6G.
\end{IEEEkeywords}

% \section*{Acronyms}
% \acuseall
% \setlength{\mylabelwidth}{0.2\columnwidth}
% \IEEEprintacronyms
% \tableofcontents
%% *************************************************************************

\section{Introduction}
\label{sec:Introduction}

\acresetall

\IEEEPARstart{R}{ecently}, \ac{iot} has experienced rapid growth, with the global number of \ac{iot} devices reaching 13.2 billion in 2022. Meanwhile, it is expected to exceed 34.7 billion by 2028 as traditional industries undergo intelligent transformations~\cite{iot_development}. \Ac{iot} has revolutionized connectivity, allowing all things to be interconnected intelligently through advanced technologies like \ac{ai}. Wireless communication technologies such as \ac{rfid}, Zigbee, cellular mobile communications, and others can provide connectivity services for \ac{iot} applications.

Specially, cellular-based \ac{iot} networks gained widespread acceptance around the world with \acs{2g} and \acs{3g} connectivity two decades ago. Nowadays, the introduction of \acs{4g} and \acs{5g} cellular networks provides benefits such as larger bandwidth, lower latency, and increased capacity, making them preferred choices for emerging IoT applications~\cite{4GM2M,Xiong2020}. For example, \ac{nbiot} based on \acs{4g} networks effectively meets the requirements of various low data speed \ac{iot} applications~\cite{NB-IoT}. 

%Existing \ac{iot} connectivity technologies, however, fall short of meeting the future demands of massive \ac{iot}, especially in terms of ultra-low power consumption. There is an urgent need for technological advancements of cellular-based \ac{iot} networks towards \acs{6g}, which is expected to offer enhanced and extended \acs{5g} capabilities in terms of achievable data rates, latency, and so on~\cite{Tataria2021}.

Unlike 4G, 5G introduces significant changes in both architecture and capabilities, particularly in its support for massive IoT deployments. In addition to NB-IoT and LTE-M technologies, 5G RedCap (Reduced Capability) was developed to cater to various IoT applications with features such as low RF complexity and cost-efficiency. However, these existing IoT technologies are insufficient to meet the future demands of massive IoT, especially regarding ultra-low power consumption. There is an urgent need for advancements in cellular-based IoT networks as we move toward 6G~{\protect\cite{6GJourney}}, which is expected to offer enhanced and extended 5G capabilities in terms of achievable data rates, latency, and so on~{\protect\cite{Tataria2021}}. The 6G vision aims to establish the foundation for a seamless cyber-physical world, requiring significant technological advancements, particularly in IoT~{\protect\cite{6GPerspective}}.

Compared to 5G, 6G technology promises to meet more stringent IoT requirements, including massive connectivity, Ultra-Reliable Low-Latency IoT Communications, improved communication protocols, extended IoT network coverage, and smart IoT devices~\cite{6GIoT}. Consequently, powering billions of IoT devices becomes one of significant challenges in 6G IoT. Traditional solutions, such as deploying power cables or regularly replacing/charging batteries, are impractical at this scale. In contrast, ambient communications utilizing backscatter modulation technology have the potential to establish long-term, maintenance-free, and energy-efficient networks due to the ultra-low power consumption characteristics~\cite{Liu2013}. As a result, both industry and academia are increasingly focusing on IoT networks based on ambient communications, also known as Ambient IoT (A-IoT). A-IoT refers to an ecosystem of numerous interconnected objects, each linked to a network through low-cost devices that utilize backscatter communications. A-IoT devices typically harvest energy from their surroundings, such as \ac{rf} signals, solar energy, vibration, or heat. A-IoT is not intended to replace existing 5G IoT technologies but is designed to extend the capabilities of the 6G system in areas where 5G falls short.

%Providing power to billions of \ac{iot} devices is a significant challenge. Traditional solutions, such as deploying power cables or replacing/charging batteries on a regular basis, are impractical. In contrast, ambient communications utilizing  backscatter modulation technology have the potential to establish long-term, maintenance-free, and energy-efficient networks due to the ultra-low power consumption characteristics~\cite{Liu2013}. Consequently, both industry and academia have now focused on \ac{iot} networks based on ambient communications, also known as the \ac{aiot}.  It is a concept referring to an ecosystem comprising numerous interconnected objects, each linked to a network through low-cost devices using backscatter communications. The \ac{aiot} devices are usually capable of harvesting energy from their surroundings, such as \ac{rf} signals, solar energy, vibration, heat, and so on~\cite{Chiu2022}. In some special cases, they may also include a small capacity battery.

%Although the 3GPP primarily focuses on RF signals in A-IoT studies, devices powered by other energy sources are not excluded in this study.  In some special cases, they may also include a small capacity battery.

Recently there are numerous studies on \ac{aiot} such as ambient backscatter modulation and so on~\cite{Yang2018,Vougioukas2019, Liu2021}. However, the majority of these studies solely focused on individual technique without taking into account the overall design and operations from the system view~\cite{Huynh2018}. Fortunately, the \ac{3gpp} has recently begun to pay attentions to \ac{aiot}, and has initiated systematic discussions and standardisation efforts in related technologies~\cite{TR_38_848, IoTMagazine}.  The efforts will undoubtedly accelerate the development of \ac{aiot} systems related to cellular mobile communications. It should be noted that, however, the studies and standardisation of key technologies in \ac{aiot} systems will face a lot of challenges and issues. This paper aims to analyze and discuss them in an exploratory manner. The main contributions of this paper are as follows:

\begin{enumerate}
\item   We thoroughly identify and examine the current state of \ac{aiot} systems within the technical framework of the 3GPP.  The analysis encompasses user cases, network topologies adopted by 3GPP for integrating backscatter communications into \ac{aiot} systems. The aim is to provide a comprehensive understanding of the existing standardization and open the door for more in-depth research.

\item  A detailed exploration is conducted on the traits of key techniques in \ac{aiot} systems. It involves the ones that are essential to the operation of \ac{aiot} systems, such as backscatter modulation, interference cancellation/avoidance, and multiple access methods.

\item We design and implement an experimental platform tailored explicitly for \ac{aiot}systems. This platform can serve as a practical demonstration, showcasing the feasibility and effectiveness of the \ac{aiot} techniques. Particularly, field experimental results are presented in terms of error rate performances under typical indoor environments.

\item This paper also tentatively presents the challenges related to system operation and performance enhancement that AIoT systems need to tackle in the next steps.

\end{enumerate}

\begin{table*}[!t]
\renewcommand{\IEEEiedlistdecl}{\setlength{\IEEElabelindent}{0pt}}
\centering
\caption{Representative Use Cases in \acs*{3gpp} \acs*{ran}}
\label{cases}
\begin{tblr}{
    width = \linewidth,
    colspec = {X[1,c,M] *4{X[2,l,t]}},
    hlines,
    hline{2} = {1}{-}{},
    hline{2} = {2}{-}{},
    vline{2-5},
    row{1} = {c,m,font=\bfseries},
    column{1} = {font=\bfseries},
    measure=vbox,
}
    \diagbox[width=\linewidth+\leftsep+\rightsep]{Group A}{Group B} & Inventory & Sensor & Positioning & Command \\
    Indoor &
    \ac{ruc} \#1: \begin{itemize}
        \item Automated warehousing
        \item Automobile manufacturing
        \item Smart laundry
        \item Same with \ac{ruc}5
    \end{itemize} &
    \ac{ruc} \#2: \begin{itemize}
        \item Smart homes
        \item Base station machine room environmental supervision
        \item Smart agriculture
    \end{itemize} &
    \ac{ruc} \#3: \begin{itemize}
        \item Ranging in a home
        \item Positioning in shopping centre
        \item Museum Guide
        \item Same with \ac{ruc}7
    \end{itemize} &
    \ac{ruc} \#4: \begin{itemize}
        \item Device Permanent Deactivation
        \item Electronic shelf label
        \item Same with \ac{ruc}8
    \end{itemize} \\
    Outdoor &
    \ac{ruc} \#5: \begin{itemize}
        \item Medical instruments inventory management and positioning
        \item Non-public network for logistics
        \item Airport terminal/shipping port
        \item Automated supply chain distribution
    \end{itemize} &
    \ac{ruc} \#6: \begin{itemize}
        \item Smart grids
        \item Forest Fire Monitoring
        \item Dairy farming
        \item Smart bridge health monitoring
    \end{itemize} &
    \ac{ruc} \#7: \begin{itemize}
        \item Finding remote lost item
        \item Location service
        \item Personal belongings finding
    \end{itemize} &
    \ac{ruc} \#8: \begin{itemize}
        \item Online modification of medical instruments status
        \item Device activation and deactivation
        \item Elderly Health Care
        \item Controller in smart agriculture
    \end{itemize}
\end{tblr}
\end{table*}

This paper is organized as follows. A detailed description on current progress of \ac{3gpp} standardization on \ac{aiot} is provided in Section~\ref{sec:3gpp}. Next, we discuss key techniques conceived for \ac{aiot} systems in Section~\ref{sec:Technologies}. We then present our demonstration system and its experimental results in Section~\ref{sec:Results}, and also discuss some open issues in Section~\ref{sec:Challenges}. Finally, our conclusions are given in Section~\ref{sec:Conclusion}.

\section{Ambient \acs*{iot} Standardization in \acs*{3gpp}}
\label{sec:3gpp}

The motivation behind the \ac{aiot} study in \ac{3gpp} is to facilitate the use of ultra-low cost and ultra-low power devices for \ac{iot} applications. Currently, study item on \ac{aiot} in \ac{3gpp} has been finished. This study specifically concentrates on \ac{iot} technologies, which is suitable for deployment in \ac{6g}. An agreement has been reached among companies regarding representative Use Cases (rUCs), connectivity topologies and deployment scenarios as outlined in~\cite{TR_38_848}. However, there are still numerous open issues that need to be solved in the stage of work item in \ac{3gpp}~\cite{TR_38_769}.

\subsection{Representative Use Cases}

To facilitate requirement definitions, \ac{3gpp} has established two sets or levels of grouping use cases of \ac{aiot} systems~\cite{TR_38_848}, i.e.,
\begin{itemize}
\item \textbf{Grouping A:} This grouping categorizes use cases based on the deployment environment, distinguishing between \textit{indoor} and \textit{outdoor} scenarios.

\item \textbf{Grouping B:} This grouping classifies use cases based on functionality and application, including \textit{inventory} management, \textit{sensor} applications, \textit{positioning}, and \textit{command} applications.
\end{itemize}
These two groupings have led to the eight kinds of \acp{ruc}, as outlined in Table~\ref{cases}. These \acp{ruc} serve as examples representing different application scenarios of \ac{aiot}. %\textcolor{myblue}{没太看懂上下文逻辑？A framework for understanding and analyzing specific use cases is provided in terms of their characteristics, requirements, and potential implementation %approaches.}

The deployment scenarios for \ac{aiot} \acp{ruc} are expected to be represented by a set of characteristics, including environment, base station characteristics, connectivity topology, spectrum, co-existence with existing \ac{3gpp} systems, and traffic assumption. As a result, such grouping results may facilitate the \ac{ran} design and implementation solutions that are tailored to the specific requirements and characteristics of a specific deployment scenario.

On the other hand, the practical deployment performance of specific user cases is heavily dependent on the capability of \ac{aiot} devices, which are quite different from traditional \ac{ue}. As a result, device categorization needs to be defined based on corresponding characteristics such as energy storage capacity and the ability to generate \ac{rf} signals for transmission.  There might be three device categories for \ac{aiot} applications, i.e.,
\begin{itemize}
\item \textbf{Device A:} No energy storage, and no independent signal generation/amplification, i.e., backscattering transmission.

\item \textbf{Device B:} Having energy storage, and no independent signal generation, i.e., backscattering transmission. The amplification for reflected signals can be included based on the stored energy.

\item \textbf{Device C:} Having energy storage and independent signal generation, i.e., active \ac{rf} components for transmission.
\end{itemize}

All the devices are expected to have low power consumption. The capabilities of different device types can vary significantly, especially between Device A and Device C. Typically, Device A or B can be powered by an \ac{rf} energy harvester at the micro-watt (\myunit{\murm W}) or sub-micro-watt (sub-\myunit{\murm W}) level, allowing for a transmission range of a few tens of meters. However, Device C may not be suitable for \ac{rf} energy as a power supply. One critical issue is that the output power of the \ac{rf} energy harvester may be even lower than the self-discharging power of the super capacitor used by Device C, which is typically a few micro-watts (\myunit{\murm W}) or higher.

Since the uplink transmission is based on backscattering, Devices A and B may not be suitable for use cases involving device-originated traffic. In contrast, Device C, with the ability to generate uplink \ac{rf} signals on its own, is assumed to support both device-terminated and device-originated traffic.
 
\subsection{Connectivity Topologies}

% \begin{figure*}[!t]
%   \centering
%   \subfloat[Topology 1]{\includegraphics[width=0.85\mymultifigwidth]{Topology A_cropped}\label{fig:f1}}
%   \hfil
%   \subfloat[Topology 2]{\includegraphics[width=0.85\mymultifigwidth]{Topology B_cropped}\label{fig:f2}}
%   \\
%   \subfloat[Topology 3]{\includegraphics[width=0.85\mymultifigwidth]{Topology C_cropped}\label{fig:f3}}
%   \hfil
%   \subfloat[Topology 4]{\includegraphics[width=0.85\mymultifigwidth]{Topology D_cropped}\label{fig:f4}}
%   \caption{Typical connectivity topology for \acs*{aiot} networks defined in \acs*{3gpp}.}
%   \label{Topology}
% \end{figure*}

\begin{figure*}[!t]
    \centering
    \includegraphics[width=0.7\linewidth]{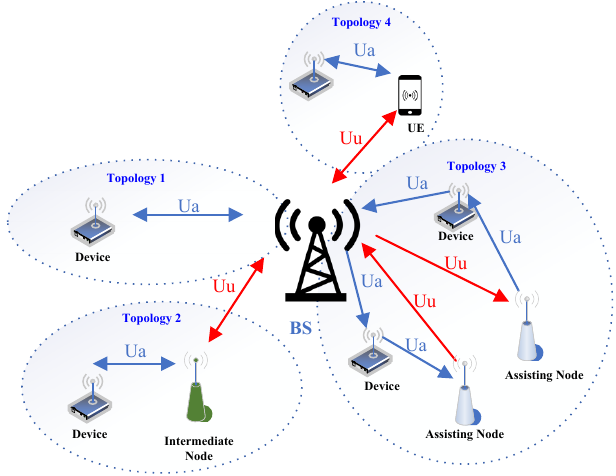}
    \caption{Typical connectivity topologies for \acs*{aiot} networks defined in \acs*{3gpp}.}
    \label{Topology}
\end{figure*}

The connectivity topologies are critical in deployment scenarios. In Fig.~\ref{Topology}, four connectivity topologies for \ac{aiot} networks have been defined and will be studied further in \ac{3gpp}. In each of these topologies, the \ac{aiot} device can receive a carrier wave, which is used to energize the device, from other nodes within or outside the topology. The links in these topologies can be either bidirectional or unidirectional, depending on the specific configuration and requirements of the network. There are two types of links in A-IoT systems, i.e., ``Uu'' and ``Ua'' links. The former denotes the air interface connecting traditional \ac{ue} to the \ac{ran}. The latter is newly designed for the connection of A-IoT devices. The nodes involved in these topologies include the \ac{bs}, assisting node, intermediate node and \ac{ue}. The placement of these nodes can be a combination of indoor and outdoor locations, allowing for flexibility in network implementation. Different topologies have distinct characteristics and suitable scenarios. Next, we briefly discuss four typical topologies.

\subsubsection{Topology 1 (\acs*{bs} $\leftrightarrow$ \acs*{aiot} Device)}

In this simplest topology, the \ac{aiot} device establishes a direct bidirectional communication link with a \ac{bs} using a new air interface, i.e., ``Ua''. This air interface is responsible for transmitting \ac{aiot} data and signalling. In those scenarios where high service availability is required over a large local or wide area, it is challenging for Device A and B to directly connect to outdoor macro- or micro-cell \acp{bs}. Device C is expected to be more suitable for ensuring continuous outdoor coverage, albeit at the expense of higher power consumption for transmission. Compared with Device A and B, Device C has ability to maintain a reliable connection with outdoor \acp{bs} while operating efficiently over a large area.

\subsubsection{Topology 2 (\acs*{bs} $\leftrightarrow$ Intermediate Node $\leftrightarrow$ \acs*{aiot} Device)}

In this topology, the bidirectional communication link between the \ac{aiot} device and the \ac{bs} is facilitated through an intermediate node. This intermediate node can be various kinds of equipments, such as a relay node, integrated access/backhaul node, \ac{ue}, repeater, and so on, only requiring limited full duplex capability. The role of the intermediate node is to transfer information between the \ac{bs} and the \ac{aiot} device, acting as a bridge in the communication process.

From the perspective of the device, this topology may have similarities to Topology 1. However, it offers the advantage of providing larger coverage. By leveraging the intermediate node, the communication range and coverage area of the \ac{aiot} system can be extended, enhancing the connectivity options.

\subsubsection{Topology 3 (\acs*{bs} $\leftrightarrow$ Assisting Node $\leftrightarrow$ \acs*{aiot} Device $\leftrightarrow$ \acs*{bs})}

This topology is specifically designed to address the use case where there is an imbalance in the communication range between the downlink and uplink. To overcome this imbalance, an assisting node is introduced to relay data between the \ac{bs} and the \ac{aiot} device. In Topology 3, the communication scenarios can vary, i.e.,
\begin{itemize}
\item The \ac{aiot} device can transmit data or signalling to the \ac{bs} while receiving data or signalling from the assisting node.

\item Alternatively, the \ac{aiot} device can receive data or signalling from the \ac{bs} while transmitting data or signalling to the assisting node.
\end{itemize}

The specific categories of the assisting node are similar with those of the intermediate node. Both the \ac{bs} and assisting node do not require full duplex capability in this topology.  On the other hand, the A-IoT device such as Device A or B has the option to receive a carrier wave from another node within or outside the topology. This flexibility allows for efficient signal energization of the A-IoT device.

\subsubsection{Topology 4 ( \acs*{ue} $\leftrightarrow$ \acs*{aiot} Device)}

In this topology, the bidirectional communication link between the \ac{aiot} device and a \ac{ue} takes place without the involvement of the \ac{bs}. The communication includes \ac{aiot} data and signalling. Moreover, it might work on an unlicensed band where the \ac{ue} does not need to establish a connection via the ``Uu'' link. However, operating in an unlicensed band imposes certain limitations, such as lower transmit power resulting in less coverage compared to licensed band. Despite these limitations, this topology is still useful, particularly for indoor environment where the coverage area is small.

\section{Potential Technologies in \acs*{aiot}}
\label{sec:Technologies}

Within the technical framework established by \ac{3gpp}, the development and maturation of various \ac{aiot} related technologies are essential prerequisites for realizing a functional \ac{aiot} system during the \ac{6g} era. Despite the emergence of diverse ambient communication technologies in recent years, this section only concentrates on potential technologies that might be applied in the \ac{aiot} system.

\subsection{Backscatter Modulation}

For simplicity, the incoming signal is defined as a sine wave with the carrier of $f_\text{in}$ and received at the \ac{aiot} device, i.e.,
\begin{align}
\label{insignal}
S_\text{in}(t)=A_\text{in}\exp\left[j\left( 2\pi f_\text{in}t+\theta_\text{in}\right) \right].
\end{align}
\noindent where $A_\text{in}$ and $\theta_\text{in}$ are the amplitude and phase of sine wave, respectively.
Afterwards, the received signal is backscattered at the antenna of the \ac{aiot} device, resulting in the product of the received signal and the backscattering coefficient $\Gamma (t)$, i.e.,
\begin{align}
\label{outsignal}
S_\text{out}(t)&=\Gamma(t) S_\text{in}(t) \notag \\
&=|\Gamma (t)| A_\text{in}\exp\left\{ j\left[ 2\pi f_\text{in}t+\theta_\text{in} +\theta_\Gamma (t) \right] \right\}.
\end{align}
\noindent where $\Gamma (t)=|\Gamma (t)|e^{j \theta_\Gamma (t)} $.  As depicted in~\eqref{outsignal}, the desired modulation scheme can be achieved by adjusting the amplitude, and phase of the backscattering coefficient. The fundamental concept behind backscatter modulation is to manipulate the antenna impedance $Z_a$ at the device, thereby generating the backscattering coefficient $\Gamma$, i.e.,
\begin{align}
\label{Gamma}
\Gamma (t) =\begin{cases}
\dfrac{Z_m (t)-Z_a^*}{Z_m(t)+Z_a}, & m=1,2,\cdots,M,\\
0, & m=0,
\end{cases}
\end{align}
where $Z_m(t)$ is the load impedance and $*$ is the complex conjugate operator. As shown in~\eqref{Gamma}, there are two modes of the device, i.e., active state ($m=1,2,\cdots,M$) and inactive state ($m=0$). In the active mode, there are $M$ switch states available. Consequently, various backscatter modulation techniques have been developed, including the \ac{ask}, \ac{fsk}, \ac{psk}, and \ac{qam}. However, the two-state modulation is commonly employed due to its simplicity. For instance, the \ac{ook} and \ac{bpsk} modulation are widely used by \ac{rfid} tags, serving as the simplest modulation schemes~\cite{rfidcoding}.

The modulated signals backscattered from the \ac{aiot} device need to be reliably detected at the receiver. In the existing literature, a few detection techniques have been already developed~\cite{Tao2019,Chen2024}. Owing to the feature of simplicity, non-coherent detection, such as the envelop detector, is the most widely used. However, it is only appropriate for \ac{ask} and \ac{fsk} modulations with low date rate that do not require carrier phase information. By contrast, in those use cases where high data rates are required, coherent detection with \ac{psk} modulation is preferred, but at the cost of pilot overhead and implementation complexity.

It is highly likely that BPSK, QPSK, and even 16QAM may be adopted as optional modulation schemes in A-IoT systems. Nevertheless, given the requirement for large transmission distances in most of rUCs, either BPSK or QPSK is preferred to be used with the coherent detection.

\subsection{Interference Cancellation/Avoidance}
 
The performance of an \ac{aiot} system is often limited by short communication ranges and low data rates. These limitations arise from the presence of \ac{dpi} and weak backscatter signals. In a traditional backscatter system, a continuous unmodulated sinusoidal wave is typically employed as the source signal, which can be easily identified by a receiver. Consequently, the non-coherent or coherent receivers can be conveniently implemented for detecting the modulated backscatter signals.

However, these detection techniques cannot be directly applied in \ac{aiot} systems. This is because these systems typically employ an unknown modulated signal from a specific legal system as the source signal, which may introduce strong \ac{dpi} to the backscatter signal. As a result, new techniques and strategies need to be developed for \ac{aiot} systems. Specially, \ac{dpi} cancellation or avoidance methods have garnered attention in 3GPP and some of them may be feasible in typical deployment scenarios, i.e.,
\begin{enumerate}
\item \textbf{Interference Cancellation:} To mitigate the \ac{dpi}, a technique such as \ac{sic} can be employed~\cite{Yang2018}. Specifically, \ac{sic} involves jointly decoding the ambient and backscatter signals when the \ac{csi} is known at the receiver. By leveraging the knowledge of the \ac{csi}, the receiver first detects the RF-source signal, then subtracts its resultant direct-link interference from the received signal, and recovers the ambient backscatter signal. Finally, based on the recovered ambient backscatter signal, the receiver re-estimates the RF-source signal. This technique enables more accurate and reliable decoding of the backscatter signals in the presence of strong \ac{dpi}.

%By leveraging the knowledge of the \ac{csi}, the \ac{dpi} component is eliminated by \ac{sic}, thereby improving the overall performance of the system. This technique enables more accurate and reliable decoding of the backscatter signals in the presence of strong \ac{dpi}.

\item \textbf{Interference Avoidance:} One approach to deal with \ac{dpi} in backscatter systems is by shifting the frequency of the backscatter signal to an adjacent non-overlapping frequency band or a guard band~\cite{Vougioukas2019}. To this end, potential \ac{dpi} can be excluded through filtering, effectively reducing its impact on the backscatter signal. This frequency shift technique can improve the received \ac{snr} and enhance the reliability of the backscatter communications. Another strategy is to exploit spatial diversity to separate the backscatter signal from the direct signal. This allows the receiver to differentiate between the two signals, even without relying on a special ambient signal or having prior knowledge of the \ac{csi}.
\end{enumerate}

\subsection{Multiple Access}
 
\begin{table*}[!t]
\renewcommand{\IEEEiedlistdecl}{\setlength{\IEEElabelindent}{0pt}}
\centering
\caption{Comparison of Different Multiple Access Methods}
\label{comparison}
\begin{tblr}{
    width = 0.95\linewidth,
    colspec = {X[0.3,c,M]X[2,l,m]X[2,c,m]},
    hlines,
    hline{2} = {1}{-}{},
    hline{2} = {2}{-}{},
    vline{2-3},
    row{1} = {c,font=\bfseries},
    column{1} = {font=\bfseries},
    columns = {rightsep=2pt},
    measure=vbox,
}
& OFDM Modulation & Backscatter Modulation \\ % Line 1

FDMA & \begin{itemize}
    \item Simply assign subsets of OFDM subcarriers to individual users, e.g., OFDMA
    \item Few users served simultaneously
\end{itemize} & \begin{itemize}
    \item Assign different frequencies to individual devices
    \item Simple but less flexible; and limited by the available bandwidth 
\end{itemize}
\\ % Line 2

TDMA & \begin{itemize}
    \item Allow multiple users to share OFDM subcarriers by dividing the signal into different time slots
    \item Strict synchronization requirement and high overhead
\end{itemize} & \begin{itemize}
    \item Different devices are assigned different time slots either deterministically or dynamically
    \item Very popular multiplexing method for backscatter communications
\end{itemize}
\\ % Line 3

CDMA & \begin{itemize}
    \item Users share either different subcarriers or time slot by using different spreading codes
    \item Various methods to combine CDMA with OFDM, e.g., Multi-carrier CDMA; usually with high implementation complexity
\end{itemize} & \begin{itemize}
    \item Devices spread the information using different orthogonal sequences and then convey the spread information by reflecting and modulating
    \item Robust against the noises but high sensitivity to time synchronization offset
\end{itemize}
\\ % Line 4

NOMA & \begin{itemize}
    \item Multiple users can be allocated on a subcarrier at the same time by adopting NOMA with OFDM
    \item High spectral efficiency with the cost of the implementation complexity and overhead
\end{itemize} & \begin{itemize}
    \item Devices in different spatial regions are multiplexed to implement PD-NOMA
    \item Limited by the implementation feasibility and power consumption
\end{itemize}
\\ % Line 5
\end{tblr}
\end{table*}

The selection of multiple access methods has a significant impact on the reliability and efficiency of \ac{aiot} systems. The majority of current studies on backscatter communications employs centralised multiple access methods, such as \ac{tdma} and \ac{fdma}, in which a central entity allocates specific resources to multiple devices in the system. \ac{fdma} is simple but has its limitation. For instance, a \ac{bs} assigns different frequencies to all \ac{aiot} devices in a \ac{fdma} system. The devices should be able to flexibly adjust their transmission frequencies within the entire system band. However, this raises the cost of individual device. Additionally, the available system bandwidth is typically limited. Both of these factors render \ac{fdma}-based systems unsuitable for large-scale deployments. On the other hand, \ac{tdma} is a more commonly used in backscatter communication systems. Unfortunately, achieving high-precision time synchronisation among devices is a huge burden for low-cost and low-power \ac{aiot} devices. As a result, new multiple access methods need to be developed specifically in \ac{aiot} systems. For example, a \ac{cbma} scheme was proposed for backscatter communication~\cite{Nanhuan2019}. It mainly consists of a correlation-based detector and power control at the device. The former aims to reduce the negative effects caused by the asynchronous problem, while the purpose of the latter is to improve the performance against the significant power difference among devices.  Furthermore, \ac{aiot} systems are expected to support a large number of devices simultaneously. Therefore, it is necessary to investigate advanced multiple access methods with efficient concurrent transmissions. Power-domain \ac{noma} is one of the methods that can achieve this goal~\cite{Guo2018}. The A-IoT devices in different spatial regions are multiplexed to implement NOMA. Unlike conventional mobile devices that can actively adjust their transmit power, the reflection coefficients for multiplexed A-IoT devices must be set to different values to more effectively leverage the benefits of power-domain NOMA.

Different multiple access methods can be applied in either OFDM or backscatter modulation schemes, which are compared and analyzed in Table~\ref{comparison}. To conclude, striking a balance of between performance and complexity, A-IoT is likely to utilize TDMA as its primary multiple access method. To enhance support for applications with a greater number of devices or higher transmission rates, there is a need to explore optional multiple access methods, including those based on CBMA, NOMA, and so on.

%To facilitate implementation, power-domain \ac{noma} can be combined with \ac{tdma} to enhance system performance. In this \ac{tdma}-\ac{noma} system, the reflection coefficients for the multiplexed devices from different groups are set to different values to utilize the power-domain \ac{noma}. This might be one of the possible multiple access candidates for \ac{aiot} systems.

%\section{Experimental Results and Analysis}
\section{Implementation and Demonstration of a \acs*{aiot}  Prototype System}
\label{sec:Results}

\begin{figure}[!t]
    \centering
    \includegraphics[width=\mysinglefigwidth]{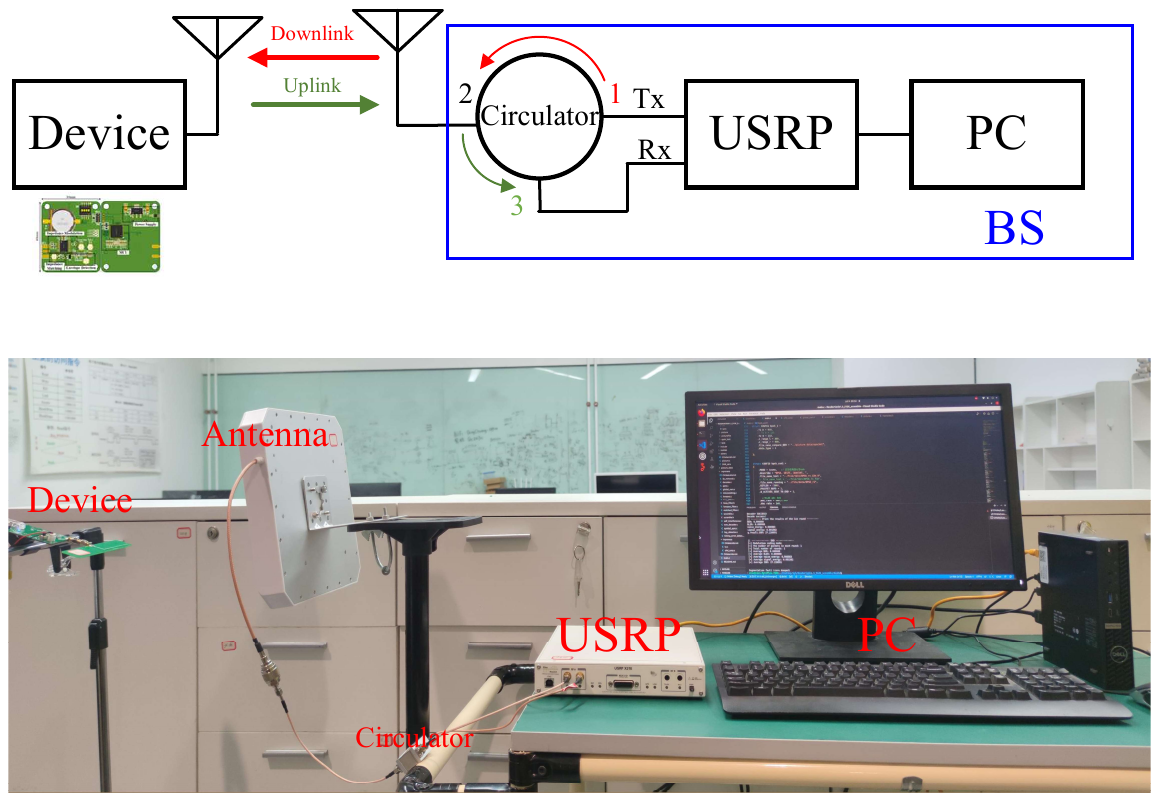}
    \caption{Illustrating of an \acs*{aiot} demonstration system.}
    \label{fig_test_bed}
\end{figure}

\begin{table}[!t]
\centering
\caption{Main Parameters of \acs*{aiot} Demonstration System}
\label{mainparameters}
\begin{tblr}{
    width = 0.9\linewidth,
    colspec = {X[2,l,m]X[1,l,m]},
    row{1} = {font=\bfseries},
    column{2} = {mode=dmath},
    cell{1,4}{2} = {mode=text},
    hlines,
    hline{2} = {1}{-}{},
    hline{2} = {2}{-}{},
    vline{2},
}
    Parameter Description & Value \\
    Carrier Frequency & 925~\myunit{MHz} \\
    Bandwidth & 1~\myunit{MHz} \\
    Modulation &  BPSK / OOK \\ 
    \ac{bs} Transmit Power & 20~\myunit{dBm} \\
    \ac{bs} Antenna Gain & 5~\myunit{dBi} \\
    Device Antenna Gain & 2~\myunit{dBi} \\
    Maximum Isolation of Circulator & 20~\myunit{dB}\\
    Insertion Loss of Circulator & 0.4~\myunit{dB}
\end{tblr}
\end{table}

%\subsection{Implementation and Demonstration of a \acs*{aiot}  System}
Although \ac{aiot} technologies are still in development, it is still very necessary to build a practical platform to demonstrate the feasibility of A-IoT systems. From a technical perspective, the backscatter communication used in future A-IoT systems will likely follow principles similar to those used in traditional RFID. We focus on showing the feasibility of using backscatter communication to achieve long-distance communication within the current 3GPP framework. As shown in Fig.~\ref{fig_test_bed}, our demonstration system comprises of a \ac{bs} and an \ac{aiot} device (also known as a tag). In this setup, only one device was considered. We plan to include more devices, which will allow us to address challenges related to network topology in the near future.The \ac{bs} consists of a general-purpose computer, a commercial \ac{usrp} X310, and a polarized panel antenna with a gain of $5~\myunit{dBi}$. The \ac{usrp} Hardware Driver library is utilized to establish concurrent transmitting and receiving threads which are responsible for various kinds of signal processing procedures. The signals interact with the \ac{usrp} X310 through Gigabit Ethernet, which handles the signal conversion between baseband and radio frequency. Meanwhile, we use a circulator\footnote{Here, the circulator model is QUEST Microwave SL8996C01, and the details of the circulator can be found at \url{http://www.questmw.com/}.} to separate the transmitted and received signals in the experiments. \textbf{Port} 1 and 3 of the circulator are connected to the \ac{usrp} X310's transmitting and receiving ports, respectively, while \textbf{Port} 2 is connected to the antenna. Finally, \ac{aiot} devices are developed by using an \acs{arm} processor and its hierarchical circuits on a four-layer printed circuit board with dimensions of $30~\myunit{mm} \times 40~\myunit{mm}$. The main parameters of the demonstration system are shown in Table~\ref{mainparameters}.

Deriving a theoretical closed-form expression that directly relates BER to distance (or SNR) is challenging due to the effects of complex fading channels in real-world environments. Therefore, we conducted field experiments to evaluate the performance of the A-IoT system in a typical indoor setting. In these experiments, we fixed the base station's location and moved the device (tag) to achieve varying communication distances. At the receiver end, specific algorithms were used to estimate different SNR values, and the corresponding BER and data rate were measured. We also implemented interference cancellation technique in our receiver to minimize the impact of interference on system performance, particularly over longer communication distances.

The communication process between the BS and the device adopts a "\textit{BS first talk, then device response}" mechanism in our demonstration system.  In \textbf{Phase 1}, the BS only sends a carrier signal to the device, which does not reflect it. Then, the power of the noise can be easily estimated by averaging the difference between the received signal and the carrier signal, i.e.,$P_n$. In \textbf{Phase 2}: the BS continues to send carrier signals, which are then reflected by the device along with its data. Subsequently, the BS receives the reflected signal.  Now, we can compute the sum of the power of the useful signal and the noise by averaging the received signal after subtracting the carrier signal, i.e.,  $P_2=P_{s}+P_n$ . Next, the  SNR can be conveniently calculated by $ \gamma=\frac{P_s}{P_n}=\frac{P_2-P_n}{P_n} $.

A customized frame structure for the data transmission is defined. Each frame consists of three parts, i.e.,  a 16-bit preamble, 992 bits of user data, and a 16-bit \ac{crc} bits. The preamble is utilized for frame synchronization and channel estimation. Then, estimated channel state information can be employed for coherent demodulation. The data part is used for transmitting the sensory data collected by the device, while the \ac{crc} bits provides data integrity verification.  After receiving the reflected signal, the BS first performs frame synchronization by cross-correlating the known preamble with the received data to search for the frame header. Once the frame header is successfully detected, the BS then utilizes the preamble for channel estimation. Subsequently, the data part undergoes a series of modules including matched filtering, self-interference cancellation, and coherent demodulation to retrieve the transmitted data from the device.

%The entire communication process between the BS and the device adopts a "\textit{BS first talk, then device response}" mechanism. In \textbf{Phase 1}, the BS only sends a carrier signal to the device, which does not reflect it. Then, the power of the noise can be easily estimated by averaging the difference between the
%received signal and the carrier signal, i.e.,$P_n$.  In \textbf{Phase 2}: the BS continues to send carrier signals, which are then reflected by the device along with its data. Subsequently, the BS receives the reflected signal.  Now, we can compute the sum of the power of the useful signal and the noise by averaging the received signal after subtracting the carrier signal, i.e.,  $P_2=P_{s}+P_n$ . Next, the  SNR can be conveniently calculated by $ \gamma=\frac{P_s}{P_n}=\frac{P_2-P_n}{P_n} $.
 
\begin{figure}[!t]
    % \centering
    \subfloat[Different modulation schemes.]{\includegraphics[width = 0.97\mymultifigwidth]{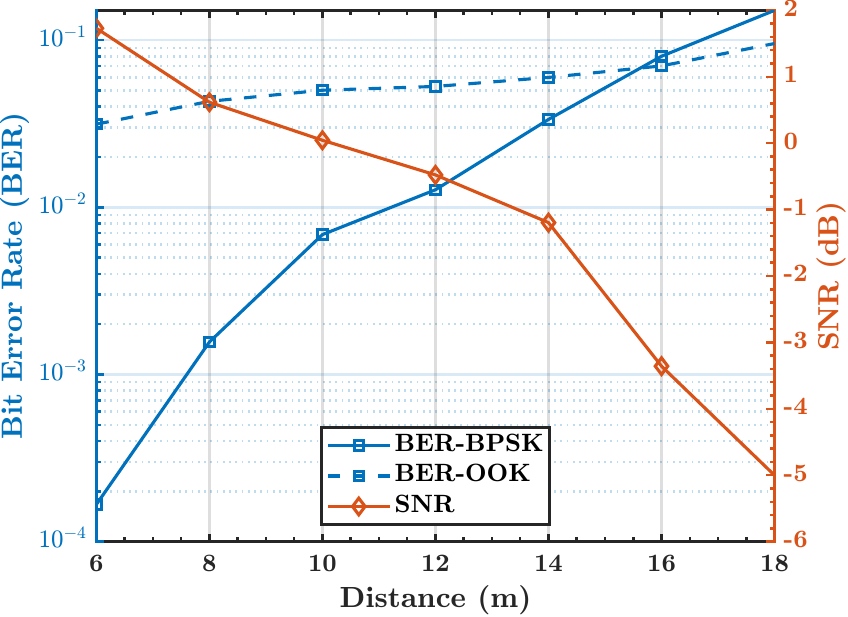}\label{fig_bpsk_ook}}
    \hfil
    \subfloat[Different data rates.]{\includegraphics[width = 0.9\mymultifigwidth]{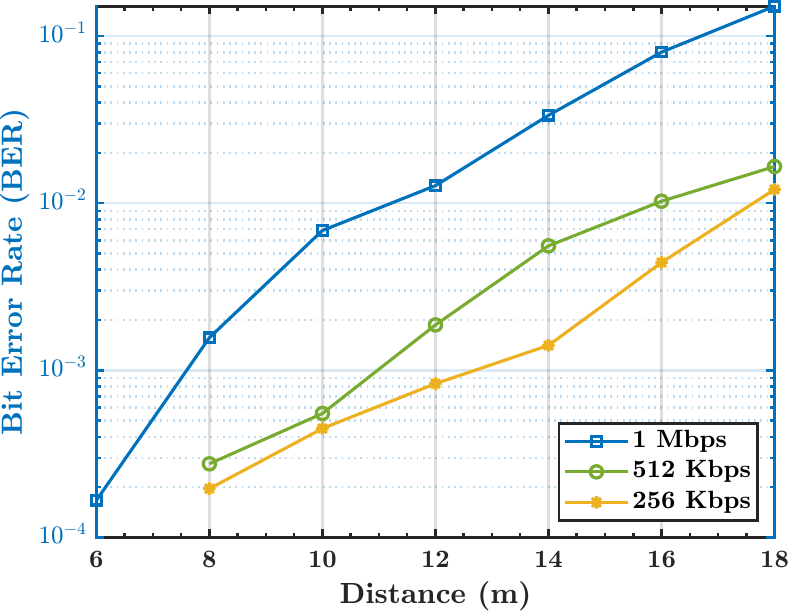}\label{fig_bpsk_tp}}
    \caption{Measured \acs*{ber} and \acs*{snr} versus the distance in an \acs*{aiot} demonstration system.}
    \label{fig_results}
\end{figure}

%\subsection{Performance Evaluation}

%Our experiments have been conducted in a typical indoor environment. The position of the \ac{bs} remains unchanged, the device (tag) can be moved to obtain a different communication distance. 
As illustrated in Fig.~\subref*{fig_bpsk_ook}, with a data rate of $1~\myunit{Mbps}$ using two modulation methods, i.e., \ac{ook} and \ac{bpsk}, respectively, the received \ac{snr} decreases as the distance rises, resulting in a higher \ac{ber}. It is obvious that the \ac{ber} of \ac{bpsk} modulation achieves the better performance. However, the BER of BPSK becomes larger than that of OOK in case of low SNR, e.g., $-5~\myunit{dB}$. This is primarily because acquiring precise phase information becomes more difficulty due to its susceptibility to the noise compared with the amplitude information. Furthermore, Fig.~\subref*{fig_bpsk_tp} indicates that the \ac{ber} performance improves when the data rate is reduced in cases of \ac{bpsk} modulation. In other words, when the required \ac{ber} performance is given, the communication distance can be extended by sacrificing the data rate in order to achieve the coverage requirement.

\section{Open Issues and Future Directions}
\label{sec:Challenges}

\subsection{Multi-Antenna Transmission}

One of the potential solutions to enhance the communication reliability and range of \ac{aiot} systems is to apply the multi-antenna technologies. In general, it is more convenient to deploy multiple antennas at the \ac{bs}, assisting node and intermediate node, where the coherent detector with multiple receiving antennas might be used to extend the communication distance.

Moreover, using multiple antennas on an \ac{aiot} device can provide diversity gain and thus further improve detection performance. It has the potential to increase the \ac{iot} device's backscatter signal power and energy harvest. More importantly, multi-antenna \ac{iot} devices allow for development of sophisticated signal processing solutions in \ac{aiot} systems, such as the \ac{stc}. For example, orthogonal \ac{stbc} can provide full spatial diversity with simple design and linear decoding complexity. It can be easily integrated into an \ac{aiot} system with multiple antennas at both the \ac{bs} and device. Depending on whether \ac{csi} is available, a coherent or non-coherent receiver with different levels of complexity and performance can be applied~\cite{Liu2021}. However, using multiple antennas in \ac{aiot} systems may increase power consumption and implementation complexity. This poses significant challenges, particularly for the devices, which must meet stringent energy and cost requirements.

\subsection{Multi-Node Joint Transmission}

For the topology discussed in Section~\ref{sec:3gpp}, there may be scenarios where multiple nodes simultaneously communicate with multiple devices. In this case, it is necessary to study how to design new transmission strategies to improve the effective distance of data or energy transfer. In general, the joint transmission of multiple \acp{bs} has been widely used to enhance the performance of mobile cellular systems. Naturally, the similar principle can also be applied to \ac{aiot} systems, especially for indoor scenarios. The distance between the \acp{bs} is usually only tens of meters, and the transmission power of each individual \ac{bs} is also limited. Therefore, multiple \acp{bs} around a given \ac{aiot} device can simultaneously send energy signals. This helps increase the communication distance between the device and the \ac{bs}, thereby increasing the coverage range of the downlink. However, due to the complexity and energy constraints of \ac{aiot} devices, coherent detection becomes very challenging on the downlink. On the other hand, the uplink signal from the same \ac{aiot} device can be received simultaneously at the neighboring multiple \acp{bs}, where spatial channel diversity benefits can be exploited with the joint detection technique to improve the performance. Therefore, developing a multi-node joint transmission strategy for \ac{aiot} systems is important but extremely challenging.

\subsection{Coexistence with Existing Networks}

When \ac{rf} signals are used to transmit energy, such as using dedicated continuous waves for energy transfer, inter-band/cell interference generally occurs. In such case, if the sole approach to improve the quality of the backscattering signal is to increase the transmission power, eliminating out-of-band emissions that interfere with signals from other systems will become extremely difficult. Thus, mobile network providers are unable to rapidly deploy \ac{aiot} networks in existing \acp{bs} with merely software updates. Instead, it is necessary to update or replace existing hardware in \acp{bs}. Furthermore, new signal algorithms are required to efficiently handle interference at various layers. The functionality of \ac{aiot} devices, on the other hand, which is typically limited by size and cost, is another key factor impacting the quick deployment within existing networks. As a result, achieving effective coexistence between \ac{aiot} systems and other existing wireless networks is a significant challenge from the perspective of network deployment.

\subsection{Device Localization}

Given the unique characteristics of backscatter technology and the limitations of device capabilities, determining the relative or absolute location of \ac{aiot} devices poses significant challenges.  Consequently, it is more feasible to prioritize less complex localization approaches from a point of view of practicality. As we know, backscatter channel has significantly different fading characteristics compared with conventional wireless links. Specifically, the emitter-to-tag channel is the product of the forward and backscatter channels, also known as a dyadic channel. Toward this end, it makes channel estimation and acquisition much more complicated resulting in increased burdens to obtain accurate localization. Furthermore, the \ac{aiot} device may necessitate the accumulation of sufficient energy to facilitate responses to activation signals sent by the \ac{bs}. However, since the \ac{bs} is unaware of the current energy status of device, it cannot estimate the time required for the devices to energise before responding. As a result, determining the individual contributions of propagation delay and charge delay to total delay is difficult for the network, making accurate positioning estimation based on propagation delay problematic.

\subsection{Spectrum Allocation}

A-IoT systems may operate in either TDD or FDD modes, each with distinct radio spectrum requirements. Therefore, the selection of frequency bands should remain flexible, encompassing both unlicensed and licensed options. When operating in unlicensed frequency bands, A-IoT systems must adhere to specific requirements. For instance, devices operating in the unlicensed bands must implement channel access schemes like Listen-Before-Talk (LBT), and are subject to maximum transmission power limits based on frequency regulations. For licensed frequency bands, considering the characteristics of backscatter communication and the goal of achieving long communication distances, A-IoT systems are more likely to be deployed in sub-6 GHz bands. However, finding suitable bands within the already crowded sub-6 GHz spectrum will present a significant challenge for the industry.

\subsection{Radio Resource Management}

Effectively utilizing limited radio resources to serve multiple IoT devices is another significant challenge for A-IoT systems. Given the massive number of A-IoT devices with small data load, traditional radio resource management that prioritize maximizing system throughput will no longer be suitable. Instead, resource allocation schemes that balance performance and fairness will become a key focus of future research. Moreover, since a plenty of A-IoT applications are delay-sensitive, resource management must be optimized using metrics that account for delay, such as Age-of-Information (AoI). In practical communication scenarios, the limitations of A-IoT devices in terms of complexity and power consumption make it difficult for these devices to transmit pilot signals or provide feedback on channel state information (CSI). As a result, inaccurate CSI estimation is inevitable in A-IoT systems, which will place higher demands on the design of resource management schemes.

\subsection{Security}

Two of the most important security concerns for \ac{aiot} systems are data confidentiality and authentication. It is crucial to ensure that devices respond only to legitimate requests made under authorized access and to prevent eavesdropping or interference with data. However, the limited processing capabilities and energy resources of \ac{aiot} devices make it difficult in achieving these goals. Therefore, it is advisable to offload security processing tasks with high complexity to the base station, which typically has larger computational capabilities. This approach aims to strike a balance between security and resource constraints. Moreover, it is essential to investigate lightweight cryptographic algorithms designed specifically for resource-constrained devices, e.g., the use of lightweight block ciphers or hash functions optimized for efficiency. Additionally, one possible solution for Device A may involve the use of pre-shared keys between \ac{aiot} device and BS, providing a certain level of security. As we know, security is usually about finding the balance between the required level of protection and the limitations of the system. The selection of specific security measures should align with use cases and perceived risks. With technological advancements, there may be ongoing developments in \ac{aiot} security tailored for resource-constrained systems.

\section{Conclusion}
\label{sec:Conclusion}

\ac{aiot} systems heavily rely on backscatter communication technology. More importantly, systematic study and development are required for \ac{aiot} to mature into a viable future system. First of all, this article has thoroughly examined and presented the ongoing \ac{aiot} standardization initiatives in \ac{3gpp}. Following that, the essential \ac{aiot} technologies associated to the \ac{3gpp} technical framework have been addressed. Furthermore, we have further implemented a comprehensive \ac{aiot} demonstration system. The feasibility of \ac{aiot} has been well demonstrated by field experimental results achieved by this system, e.g., it can support a communication rate of $1~\myunit{Mbps}$ at a distance of more than $10$ meter.  Numerous challenges persist regarding the network operation and performance improvement of AIoT systems. This paper has undertaken a preliminary study on these challenges, recognizing that there are more issues yet to be addressed.

% \appendices

% \section{Proof of Theorem~\ref{theo1}}
% \label{app:theo1}

% \blindtext
%% *************************************************************************
%\section*{Acknowledgment}
%\addcontentsline{toc}{section}{Acknowledgment}
%
%\blindtext
%% *************************************************************************
%% References section
%%
%% Can use a bibliography generated by BibTeX as a .bbl file.
%%
\bibliographystyle{IEEEtran}
%% Argument is your BibTeX string definitions and bibliography database(s).
\bibliography{IEEEabrv,Ref}
%%
%% <OR> Manually copy in the resultant .bbl file.
%% Set second argument of \begin to the number of references.
%% (used to reserve space for the reference number labels box)
%%
%\begin{thebibliography}{1}
%	
%	\bibitem{IEEEhowto:kopka}
%	H.~Kopka and P.~W. Daly, \emph{A Guide to {\LaTeX}}, 3rd~ed.\hskip 1em plus
%	0.5em minus 0.4em\relax Harlow, England: Addison-Wesley, 1999.
%	
%\end{thebibliography}
%%
%% *************************************************************************

% \begin{IEEEbiography}[{\includegraphics[width=1in,height=1.25in,clip,keepaspectratio]{Figure/PDF/KanZheng.jpg}}]{Kan Zheng}
% (Fellow, IEEE) is currently a full professor with Ningbo University, Zhejiang, China. He has rich experience in research and standardization of new emerging technologies. He has authored over 200 journal articles and conference papers in the field of wireless communications, vehicular networks, IoT and so on. He holds editorial board positions with several journals and also served in the organizing/TPC committees for conferences.
% \end{IEEEbiography}

%% *************************************************************************
% \section*{Acronyms}
% \acuseall
% \setlength{\mylabelwidth}{0.2\columnwidth}
% \IEEEprintacronyms
%% End All

\end{document}